\documentclass{ETFjee}%
\usepackage{amsfonts}
\usepackage{amsmath}
\usepackage{amssymb}
\usepackage{txfonts}
\usepackage{graphicx}%
\newtheorem{theorem}{Theorem}

\begin{document}

\title{RECONSTRUCION OF SPARSE AND NONSPARSE SIGNALS FROM A REDUCED SET OF SAMPLES}
\author{Ljubi\v{s}a Stankovi\'{c}, Isidora Stankovi\'{c}}
\thanks{This paper contains some results published in the book L. Stankovi\'{c},
"\textit{Digital signal Processing with Selected Topics}", CreateSpace,
Amazon, 2015, adapted for this publication by I. Stankovi\'{c}.}\thanks{Prof. Ljubi\v{s}a Stankovi\'{c}, University of Montenegro, ljubisa@ac.me}\thanks{MSc Isidora Stankovi\'{c}, Diploma of Imperial College (DIC) London.}
\maketitle

\begin{abstract}
Signals sparse in a transformation domain can be recovered from a reduced set
of randomly positioned samples by using compressive sensing algorithms. Simple
reconstruction algorithms are presented in the first part of the paper. The
missing samples manifest themselves as a noise in this reconstruction. Once
the reconstruction conditions for a sparse signal are met and the
reconstruction is achieved, the noise due to missing samples does not
influence the results in a direct way. It influences the possibility to
recover a signal only. Additive input noise will remain in the resulting
reconstructed signal. The accuracy of the recovery results is related to the
additive input noise. Simple derivation of this relation is presented. If a
reconstruction algorithm for a sparse signal is used in the reconstruction of
a nonsparse signal then the noise due to missing samples will remain and
behave as an additive input noise. An exact relation for the mean square error
of this error is derived for the partial DFT matrix case in this paper and
presented in form of a theorem. It takes into account very important fact that
if all samples are available then the error will be zero, for both sparse and
nonsparse recovered signals. Theory is illustrated and checked on statistical examples.

\end{abstract}

\HeaderOdd{L. Stankovi\'{c}, I. Stankovi\'{c}: Reconstruction  of Sparse and Nonsparse Signals...}
\HeaderEven{Vol. 21, No. 1, December 2015.}

\section{Introduction}

A signal can be transformed from one domain into another in various ways. Some
signals that cover the whole considered interval in one domain (where signals
are dense in that domain) could be located within much smaller regions in
another domain. We say that signals are sparse in a transformation domain if
the number of nonzero coefficients is much smaller that the total number of
signal samples. For example, a sum of discrete-time complex sinusoidal
signals, with a number of components being much lower than the number of
signal samples in the time domain, is a sparse signal in the discrete Fourier
transform (DFT) domain.

Sparse signals could be reconstructed from much fewer samples than the
sampling theorem requires. Compressive sensing is a field dealing with the
problem of signal recovery from a reduced set of samples \cite{donoho2006}%
-\cite{IsiMag}.\ As a study case, in this paper we will consider signals that
are sparse in the Fourier transform domain. Signal sparsity in the discrete
Fourier domain imposes some restrictions on the signal. Reducing the number of
samples in the analysis manifests as a noise, whose properties are studied in
\cite{16a} and used in \cite{Srdjan} to define a reconstruction algorithm. The
input noise influence is also an important topic in this analysis since the
reduced number of available samples could increase the sensitivity of the
recovery results to this noise \cite{MP2,Srdjan,SrdjanKnjiga}.
Additive noise will remain in the resulting transform.

However, if a reconstruction algorithm for a sparse signal is used in the
reconstruction of nonsparse signal then the noise, due to missing samples,
will remain and behave as an additive input noise. A relation for the mean
square error of this error is derived for the partial DFT matrix case. It
takes into account very important fact that if all samples are available then
the error will be zero, for both sparse and nonsparse recovered signals.
Theory is illustrated and checked on statistical examples.

The paper is organised as follows: after the introduction part in Section 1,
the definition of sparsity is presented in Section 2. In Section 3, the
reconstruction algorithm is presented for both one step reconstruction and the
iterative way. Also in Section 3, the analysis of the influence of additive
noise will be expanded. The reconstruction of nonsparse signals with additive
noise is shown in Section 4. In the appendix the conditions in which the
reconstruction of sparse signals is possible in general are presented.

\section{Sparsity and Reduced Set of Samples/Observations}

\begin{figure}
[ptb]
\begin{center}
\includegraphics[
height=3.2361in,
width=4.0127in
]%
{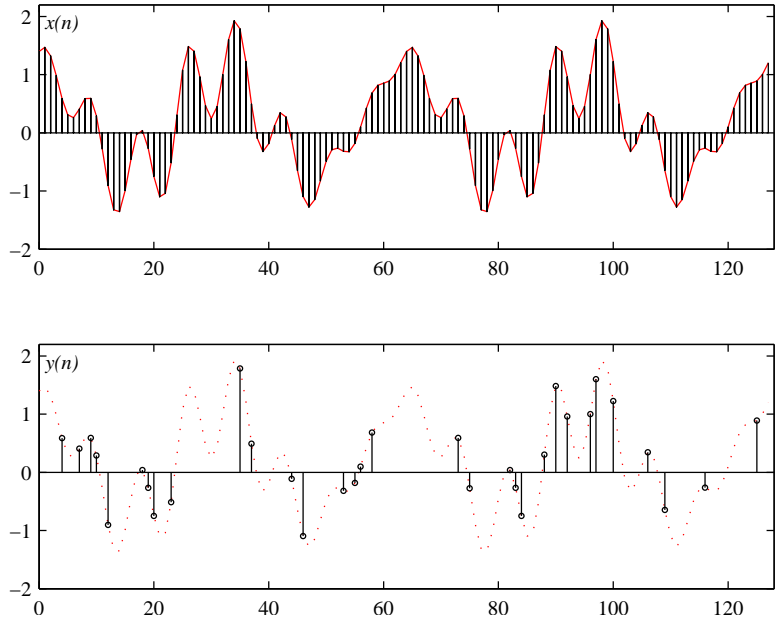}%
\caption{Signal $x(n)$ and available samples $y(n)$.}%
\label{posterslika}%
\end{center}
\end{figure}

Consider a signal $x(n)$ and its transformation domain coefficients $X(k)$,%
\index{Sparsity}
\[
x(n)=\sum_{k=0}^{N-1}X(k)\psi_{k}(n)
\]
or
\[
\mathbf{x}\mathbf{=\Psi X,}%
\]
where $\mathbf{\Psi}$ is the transformation matrix with elements $\psi_{k}%
(n)$, $\mathbf{x}$ is the signal vector column, and $\mathbf{X}$ is the
transformation coefficients vector column. For the DFT $\psi_{k}(n)=\exp(j2\pi
nk)/N$. A signal is sparse in the transformation domain if the number of
nonzero transform coefficients $K$ is much lower than the number of the
original signal samples $N$, Fig. \ref{posterslika}, i.e., if
\[
X(k)=0
\]
for%
\[
k\notin\{k_{1},k_{2},...,k_{K}\}=\mathbf{K,}%
\]
The number of nonzero samples is%
\[
\left\Vert \mathbf{X}\right\Vert _{0}=\mathrm{card}\left\{  \mathbf{X}%
\right\}  =K,
\]
where
\[
\left\Vert \mathbf{X}\right\Vert _{0}=\sum_{k=0}^{N-1}\left\vert
X(k)\right\vert ^{0}%
\]
and $\mathrm{card}\left\{  \mathbf{X}\right\}  $ is the notation for the
number of nonzero transformation coefficients in $\mathbf{X}$. Counting the
nonzero coefficients in a signal representation can be achieved by using the
so called $\ell_{0}$-norm denoted by $\left\Vert \mathbf{X}\right\Vert _{0}$.
This form is referred to as the $\ell_{0}$-norm (norm-zero) although it does
not satisfy norm properties. By definition $\left\vert X(k)\right\vert ^{0}=0$
for $\left\vert X(k)\right\vert =0$ and $\left\vert X(k)\right\vert ^{0}=1$
$\ $for $\left\vert X(k)\right\vert \neq0$.%
\index{Norm-zero}

A signal $x(n)$, whose transformation coefficients are $X(k)$, is sparse in
this transformation domain if%
\[
\mathrm{card}\left\{  \mathbf{X}\right\}  =K\ll N.
\]
For linear signal transforms the signal can be written as a linear combination
of the sparse domain coefficients $X(k)$
\begin{equation}
x(n)=\sum_{k\in\{k_{1},k_{2},...,k_{K}\}}X(k)\psi_{k}(n). \label{MeasDFT}%
\end{equation}
A signal sample can be considered as a linear combination (measurement) of
values $X(k)$. Assume that samples of $x(n)$ are available only at a reduced
set of random positions
\[
n_{i}\in\mathbf{M=}\{n_{1},n_{2},...,n_{M}\}\mathbf{\subset N}%
=\{0,1,2,3,...,N-1\}.
\]
Here $\mathbf{N}=\{0,1,2,3,...,N-1\}$ is the set of all samples of a signal
$x(n)$ and $\mathbf{M=}\{n_{1},n_{2},...,n_{M}\}$ is its random subset with
$M$ elements, $M\leq N$. The available signal values are denoted by vector
$\mathbf{y}$, Fig.\ref{posterslika},
\[
\mathbf{y=[}x(n_{1}),~x(n_{2}),~...,x(n_{M})]^{T}.
\]
%

The available samples (measurements of a linear combination of $X(k)$) defined
by (\ref{MeasDFT}), for $n_{i}\in\mathbf{M=}\{n_{1},n_{2},...,n_{M}\}$, can be
written as a system of $M$ equations
\[
\left[
\begin{array}
[c]{c}%
x(n_{1})\\
x(n_{2})\\
...\\
x(n_{M})
\end{array}
\right]  =\left[
\begin{array}
[c]{cccc}%
\psi_{0}(n_{1}) & \psi_{1}(n_{1}) &  & \psi_{N-1}(n_{1})\\
\psi_{0}(n_{2}) & \psi_{1}(n_{2}) &  & \psi_{N-1}(n_{2})\\
... & ... &  & ...\\
\psi_{0}(n_{M}) & \psi_{1}(n_{M}) &  & \psi_{N-1}(n_{M})
\end{array}
\right]  \left[
\begin{array}
[c]{c}%
X(0)\\
X(0)\\
...\\
X(N-1)
\end{array}
\right]
\]
or
\[
\mathbf{y=AX}%
\]
where $\mathbf{A}$ is the $M\times N$ matrix of
measurements/observations/available signal samples.
\index{Measurement Matrix}

The fact that the signal is sparse with $X(k)=0$ for $k\notin\{k_{1}%
,k_{2},...,k_{K}\}=\mathbf{K}$ is not included in the measurement matrix
$\mathbf{A}$ since the positions of the nonzero values are unknown. If the
knowledge that $X(k)=0$ for $k\notin\{k_{1},k_{2},...,k_{K}\}=\mathbf{K}$ were
included then a reduced observation matrix would be obtained as
\[
\left[
\begin{array}
[c]{c}%
x(n_{1})\\
x(n_{2})\\
...\\
x(n_{M})
\end{array}
\right]  =\left[
\begin{array}
[c]{cccc}%
\psi_{k_{1}}(n_{1}) & \psi_{k_{2}}(n_{1}) &  & \psi_{k_{K}}(n_{1})\\
\psi_{k_{1}}(n_{2}) & \psi_{k_{2}}(n_{2}) &  & \psi_{k_{K}}(n_{2})\\
... & ... &  & ...\\
\psi_{k_{1}}(n_{M}) & \psi_{k_{2}}(n_{M}) &  & \psi_{k_{K}}(n_{M})
\end{array}
\right]  \left[
\begin{array}
[c]{c}%
X(k_{1})\\
X(k_{2})\\
...\\
X(k_{K})
\end{array}
\right]
\]
or
\[
\mathbf{y=A}_{K}\mathbf{X}_{K}.
\]
Matrix $\mathbf{A}_{K}$ would be formed if we knew the positions of nonzero
samples $k\in\{k_{1},k_{2},...,k_{K}\}=\mathbf{K}$. It would follow from the
measurement matrix $\mathbf{A}$ by omitting the columns corresponding to the
zero-valued coefficients $X(k)$.

Assuming that there are $K$ nonzero coefficients $X(k)$, out of the total
number of $N$ values, the total number of possible different matrices
$\mathbf{A}_{K}$ is equal to the number of combinations with $K$ out of $N$.
It is equal to $\binom{N}{K}$.

\section{Signal Reconstruction}

Although the $\ell_{0}$-norm cannot be used in the direct minimization, the
algorithms based on the assumption that some coefficients $X(k)$ are equal to
zero, and the minimization of the number of remaining nonzero coefficients
that can reconstruct sparse signal, may efficiently be used.

\subsection{Direct Combinatorial Search}

The reconstruction process can be formulated as finding the positions and the
values of $K$ nonzero coefficients $X(k)$ of a sparse signal (or all signal
$x(n)$ values) using a reduced set of signal values $x(n_{i})$,
\[
n_{i}\in\mathbf{M}=\{n_{1},n_{2},...,n_{M}\}\subset\{0,1,2,...,N-1\}
\]
such that%
\[
\min\left\Vert \mathbf{X}\right\Vert _{0}\text{ \ subject to \ \ }%
\mathbf{y=AX}%
\]
where $\left\Vert \mathbf{X}\right\Vert _{0}=\mathrm{card}\{\mathbf{X}\}=K$.
Consider a discrete-time signal $x(n)$. Signal is sparse in a transformation
domain defined by the basis functions set $\psi_{k}(n)$, $k=0,1,...,N-1$. The
number of nonzero transform coefficients $K$ is much lower than the number of
the original signal samples $N$, i.e., $X(k)=0$ for
\[
k\notin\{k_{1},k_{2},...,k_{K}\}=\mathbf{K,}%
\]
$K\ll N$. A signal%
\begin{equation}
x(n)=\sum_{k\in\{k_{1},k_{2},...,k_{K}\}}X(k)\psi_{k}(n). \label{SIG_FS_SAMM}%
\end{equation}
of sparsity $K$ can be reconstructed from $M$ samples, where $M\leq N$.\ In
the case of signal $x(n)$ which is sparse in the transformation domain there
are $K$ nonzero unknown values $X(k_{1})$, $X(k_{2})$,...,$X(k_{K})$. Other
transform coefficients $X(k)$, for $k\notin\{k_{1},k_{2},...,k_{K}%
\}=\mathbf{K}$\textbf{, }are zero-valued.

Just for the beginning assume that the transformation coefficient positions
$\{k_{1}$, $k_{2}$, ..., $k_{K}\}$ are known. Then the minimal number of
equations to find the unknown coefficients (and to calculate signal $x(n)$ for
any $n$) is $K$. The equations are written for at least $K$ time instants
$n_{i}$, $i=1,2,...,M\geq K$, where the signal is available/measured,
\begin{equation}
\sum_{k\in\mathbf{K}}X(k)\psi_{k}(n_{i})=x(n_{i})\text{, for }i=1,2,...,M\geq
K.\label{Sist_RJ}%
\end{equation}
In a matrix form this system of equations is
\begin{equation}
\mathbf{A}_{K}\mathbf{X}_{K}\mathbf{=y,}\label{SusMat}%
\end{equation}
where $\mathbf{X}_{K}$ is the vector of unknown nonzero coefficients values
(at the known positions) and $\mathbf{y}$ is the vector of available signal
samples,
\begin{gather}
\mathbf{X}_{K}=[X(k_{1})~~X(k_{2})~~...~~X(k_{K})]^{T}\\
\mathbf{y}=[x(n_{1})~~x(n_{2})~~...~~x(n_{M})]^{T}\nonumber\\
\mathbf{A}_{K}=\left[
\begin{array}
[c]{cccc}%
\psi_{k_{1}}(n_{1}) & \psi_{k_{2}}(n_{1}) & ... & \psi_{k_{K}}(n_{1})\\
\psi_{k_{1}}(n_{2}) & \psi_{k_{2}}(n_{2}) & ... & \psi_{k_{K}}(n_{2})\\
... & ... & ... & ....\\
\psi_{k_{1}}(n_{K}) & \psi_{k_{2}}(n_{K}) & ... & \psi_{k_{K}}(n_{K})
\end{array}
\right]  .\label{Martr_Sampl}%
\end{gather}
Matrix $\mathbf{A}_{K}$ is the measurements matrix $\mathbf{A}$ with the
columns corresponding to the zero-valued transform coefficients $k\notin
\{k_{1}$, $k_{2}$, ..., $k_{K}\}$ being excluded. For a given set
$\{k_{1},k_{2},...,k_{K}\}=\mathbf{K}$ the coefficients reconstruction
condition can be easily formulated as the condition that system (\ref{SusMat})
has a (unique) solution, i.e., that there are $K$ independent equations,
\[
\mathrm{rank}\left(  \mathbf{A}_{K}\right)  =K.
\]
Note that this condition does not guarantee that another set $\{k_{1}%
,k_{2},...,k_{K}\}=\mathbf{K}$ can also have a (unique) solution, for the same
set of available samples. It requires $\mathrm{rank}\left(  \mathbf{A}%
_{2K}\right)  =2K$ for any submatrix $\mathbf{A}_{2K}$ of the measurements
matrix $\mathbf{A}.$

System (\ref{Sist_RJ}) is used with $K\ll M\leq N$.\ Its solution, in the mean
squared sense, follows from the minimization of the difference of the
available signal values and the values produced by inverse transform of the
reconstructed coefficients, $\min_{X(k)}\left\{  e^{2}\right\}  $ where
\begin{gather}
e^{2}=\sum_{n\in\mathbf{M}}\left\vert y(n)-\sum_{k\in\mathbf{K}}X(k)\psi
_{k}(n)\right\vert ^{2}=\nonumber\\
=\left(  \mathbf{y-A}_{K}\mathbf{X}_{K}\right)  ^{H}\left(  \mathbf{y-A}%
_{K}\mathbf{X}_{K}\right)  =\left\Vert \mathbf{y}\right\Vert _{2}%
^{2}-2\mathbf{X}_{K}^{H}\mathbf{A}_{K}^{H}\mathbf{y+X}_{K}^{H}\mathbf{A}%
_{K}^{H}\mathbf{A}_{K}\mathbf{X}_{K} \label{e_vec_h}%
\end{gather}
or
\[
\min\left\{  \left(  \mathbf{y-A}_{K}\mathbf{X}_{K}\right)  ^{H}\left(
\mathbf{y-A}_{K}\mathbf{X}_{K}\right)  \right\}
\]
where exponent $H$ denotes the Hermitian conjugate. The derivative of $e^{2}$
over a specific coefficient $X^{\ast}(p)$, $p\in\mathbf{K}$, is
\[
\frac{\partial e^{2}}{\partial X^{\ast}(p)}=2\left[  \sum_{n\in\mathbf{M}%
}y(n)-\sum_{k\in\mathbf{K}}X(k)\psi_{k}(n)\right]  \psi_{p}^{\ast}(n).
\]
The minimum of quadratic form error is reached for $\partial e^{2}/\partial
X^{\ast}(p)=0$ when
\begin{gather*}
\sum_{n\in\mathbf{M}}\psi_{p}^{\ast}(n)y(n)=\sum_{n\in\mathbf{M}}\sum
_{k\in\mathbf{K}}\psi_{k}(n)\psi_{p}^{\ast}(n)X(k)\\
\text{for }p\in\mathbf{K}\text{.}%
\end{gather*}
In matrix form this system of equations reads
\[
\mathbf{A}_{K}^{H}\mathbf{y=A}_{K}^{H}\mathbf{A}_{K}\mathbf{X}_{K}.
\]
Its solution is
\begin{equation}
\mathbf{X}_{K}\mathbf{=}\left(  \mathbf{A}_{K}^{H}\mathbf{A}_{K}\right)
^{-1}\mathbf{A}_{K}^{H}\mathbf{y.} \label{Rjesenje}%
\end{equation}
It can be obtained by a symbolic vector derivation of (\ref{e_vec_h}) as
\[
\frac{\partial e^{2}}{\partial\mathbf{X}_{K}^{H}}=-2\mathbf{A}_{K}%
^{H}\mathbf{y+2A}_{K}^{H}\mathbf{A}_{K}\mathbf{X}_{K}=0.
\]

If we do not know the positions of the nonzero values $X(k)$ for $k\in\{k_{1}%
$, $k_{2}$, ..., $k_{K}\}=\mathbf{K}$ then all possible combinations of
$\{k_{1}$, $k_{2}$, ..., $k_{K}\}\subset\mathbf{N}$ should be tested.
There\ are $\binom{N}{K}$ of them. It is not a computationally feasible
problem. Thus we must try to find a method to estimate $\{k_{1}$, $k_{2}$,
..., $k_{K}\}$ in order to recover values of $X(k)$.

\subsection{Estimation of Unknown Positions}

Solution of the minimization problem, assuming that the positions of the
nonzero signal coefficients in the sparse domain are known, is presented in
the previous subsection. The next step is to estimate the coefficient
positions, using the available samples. A simple way is to try to estimate the
positions based on signal samples that are available, ignoring unavailable
samples. This kind of transform estimate is%

\begin{equation}
\hat{X}(k)=\sum_{n\in\mathbf{M}}x(n)\varphi_{k}(n), \label{MS_SUMM}%
\end{equation}
where for the DFT $\varphi_{k}(n)=\exp(-j2\pi nk/N)$ and $n\in\mathbf{M}%
=\{n_{1},n_{2},...,n_{M}\}$. Since $\varphi_{k}(n)=N\psi_{k}^{\ast}(n)$ this
relation can be written as
\[
\mathbf{\hat{X}}=N\mathbf{\mathbf{A}}^{H}\mathbf{y}%
\]
where $\mathbf{\mathbf{A}}$ is the measurement matrix. With $K\ll M\ll N$ the
coefficients $\hat{X}(k)$, calculated with $M$ samples, are random variables.
\ Note that using (\ref{MS_SUMM}) in calculation is the same as assuming that
the values of unavailable samples $x(n)$, $n\notin\mathbf{M}$, is zero. This
kind of calculation corresponds to the result that would be achieved for the
signal transform if $\ell_{2}$-norm is used in minimization.

\textbf{Algorithm}

A simple and computationally efficient algorithm, for signal recovery, can now
be implemented as follows:

(i) Calculate the initial transform estimate $\hat{X}(k)$ by using the
available/remaining signal values
\begin{align}
\hat{X}(k)  &  =\sum_{n\in\mathbf{M}}x(n)\varphi_{k}(n)\label{MS_SUM}\\
\text{ or \ }\mathbf{\hat{X}}  &  \mathbf{=}N\mathbf{\mathbf{A}}^{H}%
\mathbf{y}\text{\textbf{.}}\nonumber
\end{align}

(ii) Set the transform values $X(k)$ to zero at all positions $k$ except the
highest ones. Alternative:

(ii) Set the transform values $X(k)$ to zero at all positions $k$ where this
initial estimate $\hat{X}(k)$ is below a threshold $T_{r}$,
\begin{align*}
X(k)  &  =0~~\text{for }k\neq k_{i}\text{, }i=1,2,...,\hat{K}\\
k_{i}  &  =\arg\{\left\vert \hat{X}(k)\right\vert >T_{r}\}.
\end{align*}
This criterion is not sensitive to $T_{r}$ as far as all nonzero positions of
the original transform are detected ($\hat{X}(k)$ is above the threshold) and
the total number $\hat{K}$ of transform values in $\hat{X}(k)$ above the
threshold is lower than the number of available samples, i.e., $K\leq\hat
{K}\leq M$.

All $\hat{K}-K$ transform values that are zero in the original signal will be
found as zero-valued.

(iii) The unknown nonzero (including $\hat{K}-K$ zero-valued) transform
coefficients could be then easily calculated by solving the set of $M$
equations for available instants $n\in\mathbf{M}$, at the detected nonzero
candidate positions $k_{i}$, $i=1,2,...,\hat{K}$,%
\begin{equation}%
{\textstyle\sum\limits_{i=1}^{\hat{K}}}
X(k_{i})\psi_{k_{i}}(n)=x(n),\text{ for }n\in\mathbf{M}\text{. }\nonumber
\end{equation}

This system of the form
\[
\mathbf{A}_{K}\mathbf{X}_{K}\mathbf{=y}%
\]
is now reduced to the problem with known positions of nonzero coefficients
(considered in the previous subsection). It is solved in the least square
sense as (\ref{Rjesenje})
\begin{equation}
\mathbf{X}_{K}=\left(  \mathbf{A}_{K}^{H}\mathbf{A}_{K}\right)  ^{-1}%
\mathbf{A}_{K}^{H}\mathbf{y.} \label{Sig_REC}%
\end{equation}
The reconstructed coefficients $X(k_{i})$, $i=1,2,...,\hat{K}$, (denoted by
vector $\mathbf{X}_{K}$) are exact, for all frequencies. If some transform
coefficients, whose true value should be zero, are included (when $K<\hat{K}$)
the resulting system will produce their correct (zero) values.

\textbf{Comments: }In general, a simple strategy can be used by assuming that
$\hat{K}=M$ and by setting to zero value only the smallest $N-M$ transform
coefficients in $\hat{X}(k)$. System (\ref{Sist_RJ}) is then a system of $M$
linear equations with $\hat{K}=M$ unknown transform values $X(k_{i})$. If the
algorithm fails to detect a component the procedure can be repeated after the
detected components are reconstructed and removed. This simple strategy is
very efficient if there is no input noise. Large $\hat{K}$, close or equal to
$M$, will increase the probability that full signal recovery is achieved in
one step. In this paper, it will be shown that in the case of an additive
(even small) input noise in all signal samples, a reduction of the number
$\hat{K}$ as close to the true signal sparsity $K$ as possible will improve
the signal to noise ratio.\ 

\textit{Example: }Consider a discrete signal
\[
x(n)=1.2e^{j2\pi n/16+j\pi/4}+1.5e^{j14\pi n/16-j\pi/3}+1.7e^{j12\pi n/16},
\]
for $0\leq n\leq15$, sparse in the DFT domain since only three DFT values are
different than zero. Assume now that its samples $x(2)$, $x(4)$, $x(11)$, and
$x(14)$ are not available. We will show that, in this case, the exact DFT
reconstruction may be achieved by: 

(i) Calculating the initial DFT estimate by
setting unavailable sample values to zero
\[
\hat{X}(k)=%
{\displaystyle\sum\limits_{n\in\mathbf{M}}}
x(n)e^{j2\pi kn/16}\mathbf{=}16\mathbf{\mathbf{A}}^{H}\mathbf{y},
\]
where
$
n\in\mathbf{M}=\{0,1,3,5,6,7,8,9,10,12,13,15\}.
$

(ii) Detecting, for example $K=3$ positions of maximal DFT values, $k_{1}$,
$k_{2}$, and $k_{3}$, and (3) calculating the reconstructed DFT values at
$k_{1}$, $k_{2}$, and $k_{3}$ from system%
\[%
\frac{1}{16}{\displaystyle\sum\limits_{i=1}^{3}}
X(k_{i})e^{j2\pi k_{i}n/16}=x(n),
\]
where $n\in\mathbf{M}=\{0,1,3,5,6,7,8,9,10,12,13,15\}$ are the instants where
the signal is available.

\begin{figure}
[ptb]
\begin{center}
\includegraphics[
height=4.7995in,
width=3.3109in
]%
{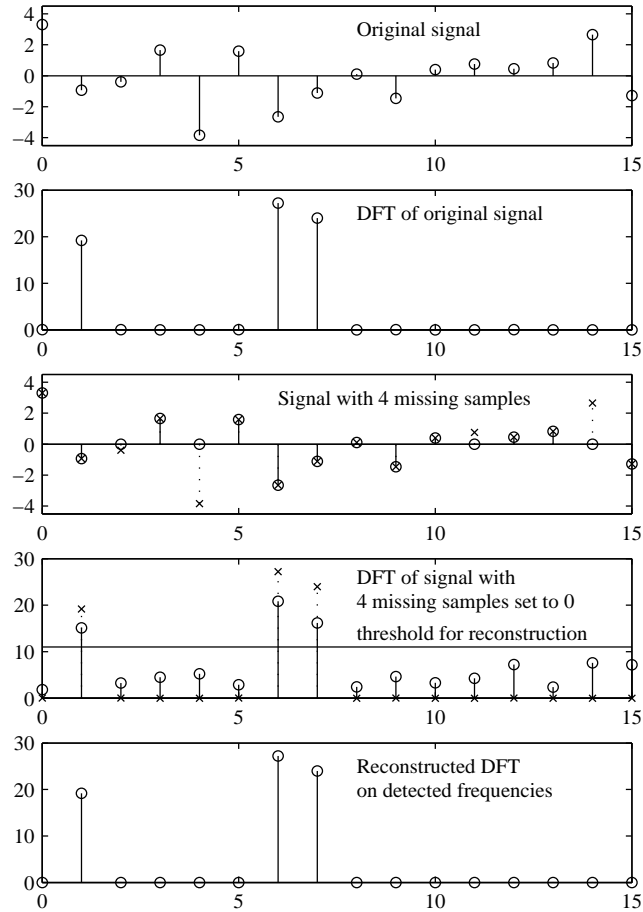}%
\caption{Original signal in the discrete-time domain (first row); the DFT of
the original signal (second row); signal with four missing samples at
$n=2,~4,11$, and $14$ set to zero (third row); the DFT of signal with missing
values being set to $0$ (fourth row). The reconstructed signal assuming that
the DFT contains components only at frequencies where the initial DFT is above
threshold (fifth row). Absolute values of the DFT and real part of signal are
shown.}%
\label{recon_expl_cs}%
\end{center}
\end{figure}

The discrete-time signal $x(n)$, with $0\leq n\leq15$ is shown in
Fig.~\ref{recon_expl_cs}. The signal is sparse in the DFT domain since only
three DFT values are different than zero (Fig.~\ref{recon_expl_cs} (second
row)). The CS signal, with missing samples $x(2)$, $x(4)$, $x(11)$, and
$x(14),$ being set to $0$ for the initial DFT estimation, is shown in
Fig.~\ref{recon_expl_cs} (third row). The DFT of the signal, with missing
values being set to $0,$ is calculated and presented in
Fig.~\ref{recon_expl_cs} (fourth row). There are three DFT values, at
$k_{1}=1$, $k_{2}=6$, and $k_{3}=7$
\[
\mathbf{K}=\{1,6,7\}
\]
above the assumed threshold, for example, at level of $11$. The rest of the
DFT values is set to $0$. This is justified by using the assumption that the
signal is sparse. Now, we form a set of equations, for these frequencies
$k_{1}=1$, $k_{2}=6$, and $k_{3}=7$ as
\[%
\frac{1}{16}{\displaystyle\sum\limits_{i=1}^{3}}
X(k_{i})e^{j2\pi k_{i}n/16}=x(n),
\]
where $n\in\mathbf{M}=\{0,1,3,5,6,7,8,9,10,12,13,15\}$ are the instants where
the signal is available.\ Since there are more equations than unknowns, the
system $\mathbf{A}_{K}\mathbf{X}_{K}\mathbf{=y}$ is solved using
$\mathbf{X}_{K}=\left(  \mathbf{A}_{K}^{H}\mathbf{A}_{K}\right)
^{-1}\mathbf{A}_{K}^{H}\mathbf{y}$. The obtained reconstructed values are
exact, for all frequencies $k$, as in Fig.~\ref{recon_expl_cs} (second row).
They are shown in Fig.~\ref{recon_expl_cs} (fifth row).

If the threshold was lower, for example at $7$, then six DFT values at
positions
\[
\mathbf{K}=\{1,6,7,12,14,15\}
\]
are above the assumed threshold. The system with six unknowns%
\[%
\frac{1}{16}{\displaystyle\sum\limits_{i=1}^{6}}
X(k_{i})e^{j2\pi k_{i}n/16}=x(n),
\]
where $n\in\mathbf{M}=\{0,1,3,5,6,7,8,9,10,12,13,15\}$ will produce the same
values for $X(1)$, $X(6)$, and $X(7)$ while the values $X(12)=X(14)=X(15)=0$
will be obtained.

If the threshold is high to include the strongest signal component only, then
the solution is obtained through an iterative procedure described in the next
subsection.

\subsection{Iterative Procedure}

If components with very different amplitudes exist and the number of available
samples is not large, then the iterative procedure should be used. This
procedure could be implemented as follows. The largest component is detected
and estimated first. It is subtracted from the signal. The next one is
detected and the signal is estimated using the frequency from this and the
previous step(s). The estimated two components are subtracted from the
original signal. The frequency of next components is detected, and the process
of estimations and subtractions is continued until the energy of the remaining
signal is negligible or bellow an expected additive noise level.\bigskip

\textbf{Algorithm}

(i) Calculate the initial transform estimate $\hat{X}_{1}(k)$ by using the
available/remaining signal values $x_{1}(n)=x(n)$
\[
\hat{X}_{1}(k)=\sum_{n\in\mathbf{M}}x(n)\varphi_{k}(n)
\]

Set the transform values $\hat{X}(k)$ to zero at all positions $k$ except the
highest one at $k=k_{1}$,

$\mathbf{\hat{K}}_{1}\mathbf{=\{}k_{1}\}$. Set the counter to $r=1.$

Form the matrix $\mathbf{A}_{1}$ using the available samples in time
$n\in\mathbf{N}_{A}$ and detected index $k\in\mathbf{\hat{K}}_{1},$ with one
nonzero component. Calculate the estimate of the transformation coefficient at
$k=k_{1}$
\[
\mathbf{\hat{X}}_{1}=\left(  \mathbf{A}_{1}^{H}\mathbf{A}_{1}\right)
^{-1}\mathbf{A}_{1}^{H}\mathbf{y.}%
\]
Calculate the signal estimation (as the inverse DFT)%
\[
\hat{x}_{1}(n)=\hat{X}_{1}(k_{1})\psi_{k_{1}}(n),\text{ for }n\in\mathbf{M}%
\]
and check
\[
\epsilon=\frac{\sum_{n\in\mathbf{M}}\left\vert x(n)-\hat{x}_{1}(n)\right\vert
^{2}}{\sum_{n\in\mathbf{M}}\left\vert x(n)\right\vert ^{2}}.
\]
If, for example $\epsilon<10^{-5}$, stop the calculation and use $x(n)=\hat
{x}_{1}(n)$. If not then go to the next step.

(ii) Set the counter to $r=r+1$. Form a signal
\[
e_{r}(n)=x(n)-\hat{x}_{r-1}(n),
\]
at the available sample positions and calculate the transform
\[
\hat{E}_{r}(k)=\sum_{n\in\mathbf{M}}e_{r}(n)\varphi_{k}(n).
\]
Set the transform values $\hat{E}_{r}(k)$ to zero at all positions $k$ except
the highest one at $k=k_{r}$. Form the set of $r$ indices, using union of the
previous maxima positions and the detected position, as
\[
\mathbf{\hat{K}}_{r}\mathbf{=\{\hat{K}}_{r-1},k_{r}\}.
\]
Form matrix $\mathbf{A}_{r}$ using the available samples in time
$n\in\mathbf{M}$ and detected $\hat{K}_{r}$ indices $k\in\mathbf{\hat{K}}_{r}%
$. Calculate the estimate of $K_{r}$ transformation coefficients
\[
\mathbf{\hat{X}}_{K_{r}}=\left(  \mathbf{A}_{r}^{H}\mathbf{A}_{r}\right)
^{-1}\mathbf{A}_{r}^{H}\mathbf{y.}%
\]
Calculate the signal%
\[
\hat{x}_{r}(n)=%
{\textstyle\sum\nolimits_{i=1}^{\hat{K}_{r}}}
\hat{X}_{r}(k_{i})\psi_{k_{i}}(n),\text{ for }n\in\mathbf{M}%
\]
and check
\[
\epsilon=\frac{\sum_{n\in\mathbf{M}}\left\vert x(n)-\hat{x}_{r}(n)\right\vert
^{2}}{\sum_{n\in\mathbf{M}}\left\vert x(n)\right\vert ^{2}}.
\]
If, for example $\epsilon<10^{-5}$, stop the calculation and use
\[
x(n)=\hat{x}_{r}(n).
\]
Else repeat step (ii).

\begin{figure}
[ptb]
\begin{center}
\includegraphics[
height=4.8285in,
width=4.0232in
]%
{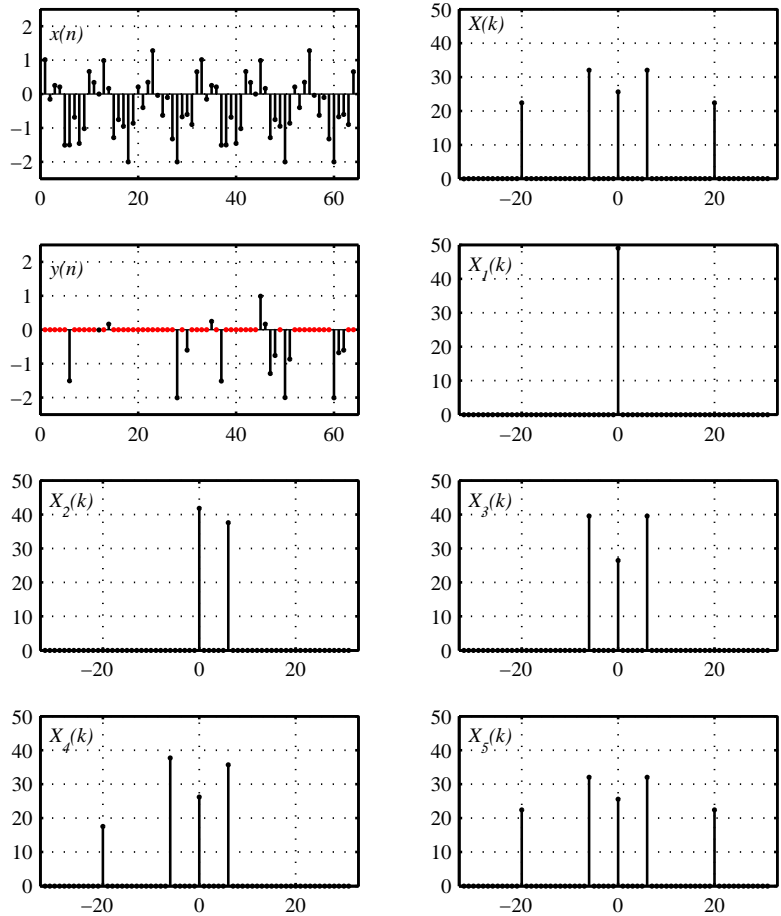}%
\caption{Iterative signal recovery}%
\label{iterativereport_knjiga}%
\end{center}
\end{figure}

\textit{Example:} Signal%
\[
x(n)=\sin(12\pi\frac{n}{N}+\frac{\pi}{4})+0.7\cos(40\pi\frac{n}{N}+\frac{\pi
}{3})-0.4
\]
with $N=64$ is shown in Fig.\ref{iterativereport_knjiga}. Small number of
samples is available $M=16$ with different signal amplitudes, making one-step
recovery impossible. The available signal samples $y(n)$ are shown in
Fig.\ref{iterativereport_knjiga} (second row, left). The iterative procedure
is used and, for the detected DFT positions during the iterations, the
recovered signal is calculated according to the presented algorithm. The
recovered DFT values in the $r$th iteration are denoted as $X_{r}(k)$ and
presented in Fig.\ref{iterativereport_knjiga}. After first iteration the
strongest component is detected and its amplitude is estimated. At this stage,
other components behave as noise and make amplitude value inaccurate. Accuracy
improves as the number of detected components increases in next iterations.
After five steps the agreement between the reconstructed signal and the
available signal samples was complete. Then the algorithm is stopped. The DFT
of the recovered signal is presented as $X_{5}(k)$ in the last subplot of
Fig.\ref{iterativereport_knjiga}. Its agreement with the DFT of the original
signal, Fig.\ref{iterativereport_knjiga} (first row, right) is complete.

\subsection{Unavailable/Missing Samples Noise}

The initial DFT calculation in reconstruction algorithms is done assuming
zero-valued missing samples. The initial calculation quality has a crucial
importance to the successful signal recovery. With a large number of randomly
positioned missing samples, the missing samples manifest as a noise in this
initial transform. Once the reconstruction conditions are met for a sparse
signal and the exact reconstruction is achieved, the noise due to missing
samples does not influence the results in a direct way. It influences the
possibility to recover a signal. The accuracy of the recovery results is
related to the additive input noise only. The input noise is transformed by
the recovery algorithm into a new noise depending on the signal sparsity and
the number of available samples. A simple analysis of this form of noise will
be presented in the second part of this section.

Consider a sparse signal in the DFT domain with nonzero coefficients $X(k)$ at
the positions $k\in\mathbf{K=}\{k_{1},k_{2},...,k_{K}\}$
\[
x(n)=\sum_{p=1}^{K}A_{p}e^{j2\pi nk_{p}/N},
\]
where $A_{p}$ are the signal component amplitudes. The initial DFT is
calculated using $n\in\mathbf{M}=\{n_{1},n_{2},...,n_{M}\}$%
\begin{equation}
X(k)=\sum_{n\in\mathbf{M}}x(n)e^{-j2\pi nk/N}=\sum_{n\in\mathbf{M}}\sum
_{p=1}^{K}A_{p}e^{-j2\pi n(k-k_{p})/N}. \label{MS_SUMMM}%
\end{equation}
We can distinguish two cases:

(1) For $k=k_{i}\in\{k_{1},k_{2},...,k_{K}\}$ then, with $M=\mathrm{card}%
(\mathbf{M})$,
\[
X(k_{i})=A_{i}M+\sum_{n\in\mathbf{M}}\sum_{\substack{p=1\\p\neq i}}^{K}%
A_{p}e^{-j2\pi n(k_{i}-k_{p})/N}.
\]
The value of
\begin{equation}
\Xi=\sum_{n\in\mathbf{M}}\sum_{\substack{p=1\\p\neq i}}^{K}A_{p}e^{-j2\pi
n(k_{i}-k_{p})/N} \label{noise_EPSEE}%
\end{equation}
with random set $\mathbf{M}=\{n_{1},n_{2},...,n_{M}\}$, for $1\ll M\ll N$, can
be considered as a random variable. Its mean over different realizations of
available samples (different realizations of sets $\mathbf{M}$) is
$\mathrm{E}\{\Xi\}=0$.The mean value of $X(k_{i})$ is
\[
\mathrm{E}\{X(k_{i})\}=A_{i}M.
\]

(2) For $k\notin\{k_{1},k_{2},...,k_{K}\}$ the mean value of (\ref{MS_SUMMM})
is
\[
\mathrm{E}\{X(k)\}=0.
\]
The mean value of (\ref{MS_SUMMM}) for any $k$ is of the form
\[
\mathrm{E}\{X(k)\}=M\sum_{p=1}^{K}A_{p}\delta(k-k_{p}),
\]
Its variance is \cite{16a, Knjiga}
\begin{equation}
\sigma_{N}^{2}(k)=\mathrm{var}(X(k))=\sum_{p=1}^{K}A_{p}^{2}M\frac{N-M}%
{N-1}\left[  1-\delta(k-k_{p})\right]  \text{.} \label{sign}%
\end{equation}
The ratio of the signal amplitude $X(k_{1})$ and the standard deviation
$\sigma_{N}(k)$ for $k\neq k_{1}$ is crucial parameter (Welsh bound for
coherence index $\mu$ of measurement matrix $\mathbf{A}$) for correct signal
detection. Its value is
\[
\frac{\sigma_{N}(k)}{\left\vert X(k_{1})\right\vert }=\sqrt{\frac{N-M}%
{M(N-1)}}\text{.}%
\]
Note that the variance in a multicomponent signal with $K>1$ is a sum of the
variances of individual components at all frequencies $k$ except at $k_{i}%
\in\{k_{1},k_{2},...,k_{K}\}$ when the values are lower for $\left\vert
A_{i}\right\vert ^{2}M\frac{N-M}{N-1}$ since all component values are added up
in phase, without random variations.

According to the central limit theorem, for $1\ll M\ll N$ the real and
imaginary parts of the DFT value for noise only positions $k\notin
\{k_{1},k_{2},...,k_{K}\}$ can be described by Gaussian distribution,
$\mathcal{N}(0,\sigma_{N}^{2}/2)$ with zero-mean and variance $\sigma_{N}%
^{2}=\sigma_{N}^{2}(k)$.\ Real and imaginary parts of the DFT value, at the
signal component position $k_{i}\in\{k_{1},k_{2},...,k_{K}\}$, can be
described by the Gaussian distributions%
\begin{align*}
&  \mathcal{N}(M\operatorname{Re}\{A_{p}\},\sigma_{S_{p}}^{2}/2),\text{ and
}\\
&  \mathcal{N}(M\operatorname{Im}\{A_{p}\},\sigma_{S_{p}}^{2}/2),
\end{align*}
respectively, where $\sigma_{S_{p}}^{2}=\sigma_{N}^{2}-A_{p}^{2}M\frac
{N-M}{N-1}$, according to (\ref{sign}).

\begin{figure}
[ptb]
\begin{center}
\includegraphics[
height=3.4105in,
width=4.1212in
]%
{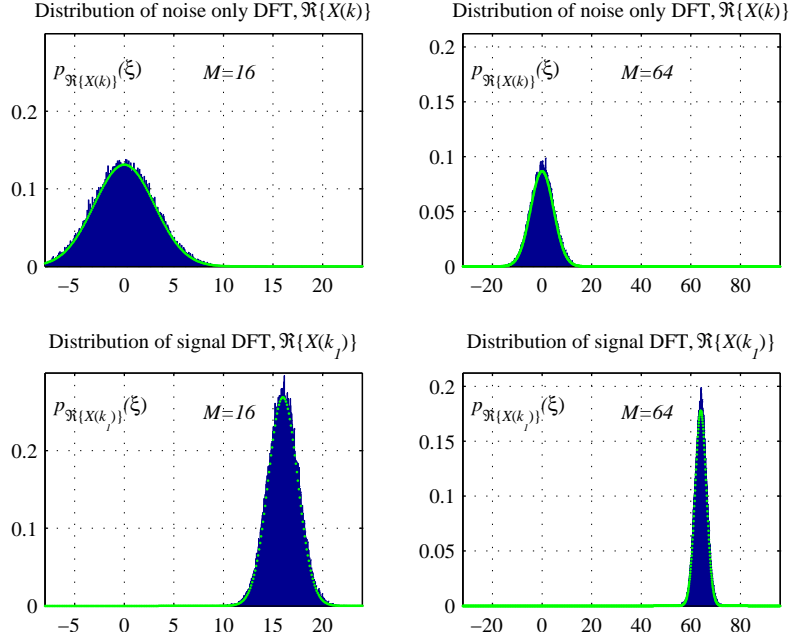}%
\caption{Histograms and Gaussian probability density functions for the signal
and noise only positions in the DFT for a three-component signal with $N=128$
and $M=16$ (left) and $M=64$ (right). The histograms are calculated in
$10^{5}$ random realizations of $M$ available samples and random signal
frequency positions.}%
\label{hist_miss_samp}%
\end{center}
\end{figure}

\textit{Example:} For a discrete-time signal
\begin{equation}
x(n)=A_{1}e^{j2\pi k_{1}n/N}+A_{2}e^{j2\pi k_{2}n/N}+A_{2}e^{j2\pi k_{3}%
n/N},\label{ssiigg}%
\end{equation}
with $N=64$, $A_{1}=1$, $A_{2}=1/2$, $A_{3}=1/4$, the DFT is calculated using
a random set of $M=16$ samples. Calculation is performed with $10^{5}$ random
realizations with randomly positioned $M$ samples and random values of
$k_{1},k_{2},$ and $k_{3}$. Histogram of the DFT values, at a noise only
position $k\notin\{k_{1},k_{2},k_{3}\}$ and at the signal component $k=k_{1}$
position, is presented in Fig.\ref{hist_miss_samp} (left). Histogram of the
DFT real part is shown, along with the corresponding Gaussian functions
$\mathcal{N}(0,\frac{21}{16}\frac{N-M}{N-1})$ and $\mathcal{N}(M,\frac{5}%
{16}\frac{N-M}{N-1})$, shown by green dots. The same calculation is repeated
with $M=64$, Fig.\ref{hist_miss_samp} (right). We can see that the mean value
of the Gaussian variable $X(k)$ can be used for the detection of the signal
component position. Also the variance is different for noise only and the
signal component positions. It can also be used for the signal position
detection. In the case with $M=16$, the histograms are close to each other,
meaning that there is a probability that a signal component is missdetected.
Histograms are well separated in the case when $M=64$. It means that the
signal component will be detected with an extremely high probability in this
case. Calculation of the detection probability is straightforward with the
assumed probability density functions.%

The spark based relation can be obtained within the framework of the previous
analysis if we assume that the noises (\ref{noise_EPSEE}) due to missing
samples coming from different components of the same (unity) amplitude $A_{i}$
are added up (equal amplitudes are the worst case for this kind of analysis)
with the same phase to produce \cite{Knjiga},
\begin{equation}
X(k)=\sum_{n\in\mathbf{M}}\sum_{p=1}^{K}e^{-j2\pi n(k-k_{p})/N}=K\sum
_{n\in\mathbf{M}}e^{-j2\pi n(k-k_{p})/N}\label{noise_EPS}%
\end{equation}
at some frequency $k\notin\{k_{1},k_{2},...,k_{K}\}$. Random variable
$\sum_{n\in\mathbf{M}}e^{-j2\pi n(k-k_{p})/N}$ (since $n\in\mathbf{M}$ is
random) should also assume its maximal possible value (calculated over all
possible $k_{p}$ and all possible positions $k$, $k\neq k_{p}$). The maximal
possible value of this variable is related to the coherence index $\mu$ of the
partial DFT matrix defined as%
\begin{equation}
\mu=\max\left\vert \mu(k,k_{p})\right\vert =\frac{1}{M}\max_{k,k_{p}%
}\left\vert \sum\limits_{n\in\mathbf{M}}e^{-j2\pi n(k-k_{p})/N}\right\vert
.\label{mimi}%
\end{equation}
It means that maximal possible value of this variable is $\mu M$. It should
also\ be assumed that $(K-1)$ remaining noise components (due to missing
samples) at the component position $k=k_{p}$ assume the same maximal value
$\mu M$ and that all of them subtract in phase from the signal mean value $M$
at $k=k_{p}$. Condition for the correct detection of a component position at
$k=k_{p}$ is then such that the minimal possible amplitude of the component
$M-M\mu(K-1)$ is greater than the maximal possible noise $M\mu K$ at
$k\notin\{k_{1},k_{2},...,k_{K}\}$, i.e.,
\[
M-M\mu(K-1)>M\mu K
\]
or
\[
K<\frac{1}{2}(1+\frac{1}{\mu})=\frac{1}{2}\mathrm{spark}(\mathbf{A}),
\]
where $\mathrm{spark}(\mathbf{A})$ is the spark of the measurement matrix
$\mathbf{A}$ (spark of matrix $\mathbf{A}$ is defined as the smallest number
of dependent columns or rows). According to many very unlikely assumptions
that has been made, we can state that this is a very pessimistic bound for
$K$. Therefore, for a high degree of randomness, a probabilistic approach may
be more suitable for the analysis than the spark based relation.

\subsection{Additive Noise Influence}

Assume an additive noise $\varepsilon(t)$ in the input signal. In a matrix
form this system of $M$ linear equations with $K$ unknowns reads%

\[
\mathbf{y+\varepsilon}=\mathbf{AX}_{K}%
\]
The solution follows for
\[
\mathbf{X}_{K}=\mathbf{X}_{KS}+\mathbf{X}_{KN}%
\]
where $\mathbf{X}_{KS}$ and $\mathbf{X}_{KN}$ are the reconstructed signal and
noise components respectively.

Assume that the reconstruction conditions are met and the positions of nonzero
coefficients $\mathbf{K=}\{k_{1}$, $k_{2}$, ..., $k_{K}\}$ can be determined
through a single step or iterative procedure \cite{noisesfacta,Srdjan, LJSIS}.
The equations to find the unknown coefficients are written for $M>K$ time
instants $n_{i}$, $i=1,2,...,M\geq K$
\begin{equation}
\mathbf{A}_{K}\mathbf{X}_{K}\mathbf{=y+\varepsilon} \label{SusMatS}%
\end{equation}
where $\mathbf{X}_{K}=[X(k_{1})~~X(k_{2})~~...~~X(k_{K})]^{T}$ is the vector
of unknown nonzero coefficients values (at the determined positions) and
$\mathbf{y}$ is the vector of the available signal samples $\mathbf{y}%
=[x(n_{1})~~x(n_{2})~~...~~x(n_{M})]^{T}.$The matrix $\mathbf{A}_{K}$ is the
measurements matrix $\mathbf{A}$ with the columns corresponding to the
zero-valued transform coefficients $k\notin\{k_{1}$, $k_{2}$, ..., $k_{K}\}$
being excluded. For a given set $\{k_{1},k_{2},...,k_{K}\}=\mathbf{K}$ the
coefficients reconstruction condition can be easily calculated as
\begin{equation}
\mathbf{X}_{K}\mathbf{=}\left(  \mathbf{A}_{K}^{H}\mathbf{A}_{K}\right)
^{-1}\mathbf{A}_{K}^{H}(\mathbf{y}+\varepsilon). \label{rec_rel}%
\end{equation}
where $\mathbf{X}_{KS}=\left(  \mathbf{A}_{K}^{H}\mathbf{A}_{K}\right)
^{-1}\mathbf{A}_{K}^{H}\mathbf{y}$ and $\mathbf{X}_{KN}=\left(  \mathbf{A}%
_{K}^{H}\mathbf{A}_{K}\right)  ^{-1}\mathbf{A}_{K}^{H}\mathbf{\varepsilon}$ is
the noise influence to the reconstructed signal coefficients.

The input signal-to-noise (SNR) ratio, if all signal samples were available, is%

\[
SNR_{i}=10\log\frac{\sum_{n=0}^{N-1}\left\vert x(n)\right\vert ^{2}}%
{\sum_{n=0}^{N-1}\left\vert \varepsilon(n)\right\vert ^{2}}=10\log\frac{E_{x}%
}{E_{\varepsilon}}.
\]

Assume the noise energy in $M$ available samples used in the reconstruction is%
\begin{equation}
E_{\varepsilon A}=\sum_{n\in\mathbf{M}}\left\vert
\varepsilon(n)\right\vert ^{2}.
\end{equation}The correct amplitude in the signal transform at the frequency $k_{p}$, in the
case if all signal samples were used, would be $NA_{p}$. To compensate the
resulting transform for the known bias in amplitude when only $M$ available
samples are used we should multiply the coefficient by $N/M$. It means that is
a full recovery, a signal transform coefficient should correspond to the
coefficient of the original signal with all signal samples being used. The
noise in the transform coefficients will also be multiplied by the same
factor. Therefore, its energy would be increased to $E_{\varepsilon A}%
N^{2}/M^{2}$. The signal-to-noise ratio in the recovered signal would be
\begin{equation}
SNR=10\log\frac{\sum_{n=0}^{N-1}\left\vert x(n)\right\vert ^{2}}%
{\frac{N^{2}}{M^{2}}\sum_{n\in\mathbf{M}}\left\vert
\varepsilon(n)\right\vert ^{2}}%
\end{equation}
If the distribution of noise in the samples used for reconstruction is the
same as in other signal samples then $\sum_{n\in\mathbf{M}}\left\vert \varepsilon(n)\right\vert ^{2}=M\sigma_{\varepsilon}^{2}$
and
\begin{gather}
SNR=10\log\frac{\sum_{n=0}^{N-1}\left\vert x(n)\right\vert ^{2}}%
{\frac{N^{2}}{M^{2}}M\sigma_{\varepsilon}^{2}}
=SNR_{i}-10\log\left(  \frac{N}{M}\right)  .
\end{gather}
Therefore, a signal reconstruction that would be based on the initial estimate
(\ref{MS_SUM}) would worsen SNR, since $N>M$.

Since only $K$ out of $N$ DFT coefficients are used in the reconstruction the
energy of the reconstruction error is reduced for the factor of $K/N$ as well.
Therefore, the energy of noise in the reconstructed signal is
\[
E_{\varepsilon R}=\frac{K}{N}\frac{N^{2}}{M^{2}}\sum_{n\in\mathbf{M}}\left\vert \varepsilon(n)\right\vert ^{2}.
\]
The output signal to noise ratio in the reconstructed signal is
\cite{Srdjan,LJSIS,Knjiga}%
\begin{equation}
SNR=10\log\frac{\sum_{n=0}^{N-1}\left\vert x(n)\right\vert ^{2}}{\frac
{KN}{M^{2}}\sum_{n\in\mathbf{M}}\left\vert \varepsilon(n)\right\vert ^{2}%
}=10\log\frac{\sum_{n=0}^{N-1}\left\vert x(n)\right\vert ^{2}}{\frac{K}{M}%
\sum_{n=0}^{N-1}\left\vert \varepsilon(n)\right\vert ^{2}}. \label{final_snr}%
\end{equation}

It is related to the input signal to noise ration $SNR_{i}$ as
\begin{equation}
SNR=SNR_{i}-10\log\left(  \frac{K}{M}\right)  . \label{out_in_relation}%
\end{equation}

\textit{Example:} Theory is illustrated on a four component noisy signal%
\begin{align*}
x(n)  &  =A_{1}\exp(j2\pi k_{1}n/N)+A_{2}\exp(j2\pi k_{2}n/N)\\
&  +A_{3}\exp(j2\pi k_{3}n/N)+A_{4}\exp(j2\pi k_{4}n/N)+\varepsilon(n)
\end{align*}
as well, where $A_{1}=1,$ $A_{2}=0.75$, $A_{3}=0.5$, $A_{4}=0.67$, $N=257,$
and $\{k_{1},k_{2},k_{3},k_{4}\}=\{58,117,21,45\}$. The signal is
reconstructed using iterative calculation to find nonzero coefficients
$\mathbf{K=}\{k_{1}$, $k_{2}$, ..., $k_{K}\}$ and (\ref{rec_rel}) to find the
signal. The results are presented in the Table \ref{tab1}.
The agreement of the numerical statistical results $SNR_{S}$ with this simple
theory in analysis of noise influence to the reconstruction of sparse signals
$SNR_{T}$ is high for all considered $SNR_{i}$.

\begin{table}%
\centering
\caption{Signal to noise ratio in the reconstructed signal according to the theory $SNR_T$ and the statistics $SNR_S$ for various $M$. Input signal to noise ratio is denoted by $SNR_i$}
\label{tab1}
\begin{tabular}
[c]{||crrrr||}\hline\hline
SNR in [dB] & $M=128$ & $M=160$ & $M=192$ & $M=224$\\
\hline
$SNR_{i}$ & 3.5360 & 3.5326 & 3.5788 & 3.5385\\
$SNR_{T}$ & 18.5953 & 19.5644 & 20.3562 & 21.0257\\
$SNR_{S}$ & 18.7203 & 19.5139 & 20.2869 & 21.7302\\\hline\hline
\end{tabular}
\end{table}

\section{Nonsparse Signal Reconstruction}

According to the results in previous section, the missing samples can be
represented by a noise influence. Assume that we use a reconstruction
algorithm for a signal of sparsity $K$ on a signal whose DFT coefficients
$\mathbf{X}$ are not sparse (or not sufficiently sparse). Denote by
$\mathbf{X}_{K}$ the sparse signal with $K$ nonzero coefficients equal to the
largest $K$ coefficients of $\mathbf{X}$. Suppose that the number of
components $K$ and the measurements matrix satisfy the reconstruction
conditions so that a reconstruction algorithm can detect (one by one or at
once) largest $K$ components ($A_{1}$, $A_{2}$,...$A_{K}$) and perform signal
reconstruction to get $\mathbf{X}_{R}$. The remaining $N-K$ components
($A_{K+1}$,$A_{K+2}$,...,$A_{N}$) will be treated as noise in these $K$
largest components. Variance of a signal component is $\left\vert
A_{i}\right\vert ^{2}M(N-M)/(N-1).$ After reconstruction the variance is
\[
\left\vert A_{i}\right\vert ^{2}\frac{N^{2}}{M^{2}}\frac{M(N-M)}{N-1}%
\cong\left\vert A_{i}\right\vert ^{2}N\frac{N-M}{M}.
\]
The total energy of noise in the reconstructed $K$ largest components
$\mathbf{X}_{R}$ will be%
\[
\left\Vert \mathbf{X}_{R}\mathbf{-X}_{K}\right\Vert _{2}^{2}=KN\frac{N-M}%
{M}\sum_{i=K+1}^{N}\left\vert A_{i}\right\vert ^{2}%
\]
Denoting the energy of remaining signal, when the $K$ largest are removed from
the original signal, by
\[
\left\Vert \mathbf{X-X}_{K}\right\Vert _{2}^{2}=N\sum_{i=K+1}^{N}\left\vert
A_{i}\right\vert ^{2}%
\]
we get
\[
\left\Vert \mathbf{X}_{R}\mathbf{-X}_{K}\right\Vert _{2}^{2}=K\frac{N-M}%
{M}\left\Vert \mathbf{X-X}_{K}\right\Vert _{2}^{2}.
\]
If the signal is sparse, i.e., $\mathbf{X=X}_{K}$, then
\[
\left\Vert \mathbf{X}_{R}\mathbf{-X}_{K}\right\Vert _{2}^{2}=0.
\]
The same result follows if $N=M$%
\[
\left\Vert \mathbf{X}_{R}\mathbf{-X}_{K}\right\Vert _{2}^{2}=0.
\]
That is, the error will be zero if a complete DFT matrix is used in
calculation.

Using Schwartz inequality $\left\Vert \mathbf{X}\right\Vert _{2}%
\leq\frac{1}{\sqrt{N}}\left\Vert \mathbf{X}\right\Vert _{1}$ follows
\[
\left\Vert \mathbf{X}_{K}\mathbf{-X}_{R}\right\Vert _{2}\leq\sqrt{\frac
{N-M}{M}\frac{K}{N-K}}\left\Vert \mathbf{X-X}_{K}\right\Vert _{1}.
\]
It means that if $\left\Vert \mathbf{X-X}_{K}\right\Vert _{1}$ is minimized
then the upper bound of the error $\left\Vert \mathbf{X}_{K}\mathbf{-X}%
_{K}\right\Vert _{2}$ is also minimized.

Based on the previous results we can easily get the following result.

\subsection{Theorem for the Error in a Nonsparse Signal Reconstruction}

\renewcommand\thetheorem{:}

\begin{theorem}
Consider a signal $x(n),\ n=0,1,...,N-1$, with transformation coefficients
$\mathbf{X}$ and unknown sparsity, including the case when the signal is not
sparse. Assume that  $M\le N$ time domain signal samples, corrupted with 
additive white noise
with variance $\sigma_{\varepsilon}^{2}$, are available.
The signal is reconstructed assuming that its sparsity is $K$.
Denote the reconstructed signal by $\mathbf{X_{R}}$, set of its nonzero 
positions by  $\mathbf{K}$, and the corresponding original signal transform by 
$\mathbf{X_{K}}$ where
${X_{K}(k)}={X}(k)$ for $k \in\mathbf{K}$ and ${X_{K}(k)%
}={0}$ for $k \notin\mathbf{K}$.
The total error in the reconstructed
signal, with respect to the original signal at the same nonzero coefficient
positions, is
\[
\left\Vert \mathbf{X}_{K}\mathbf{-X}_{R}\right\Vert _{2}^{2}=K\frac{N-M}%
{M}\left\Vert \mathbf{X-X}_{K}\right\Vert _{2}^{2}+\frac{K}{M}N\sigma
_{\varepsilon}^{2}.
\]

\textrm{\rm Proof of this theorem easily follows from the presented analysis \cite{Knjiga}.}

\end{theorem}

\begin{figure}
[ptb]
\begin{center}
\includegraphics[
height=2.1943in,
width=4.2989in
]%
{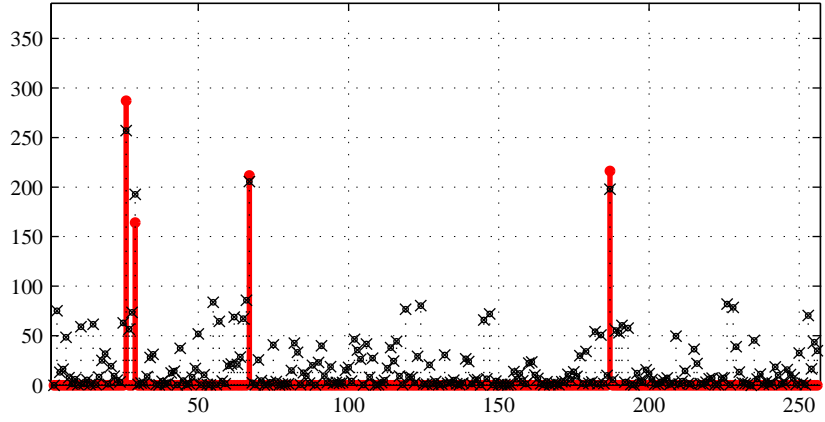}%
\caption{Single realization reconstruction of $K=4$ largest signal components
of a nonsparse noisy signal.}%
\label{input_noise_stat_snr_noise}%
\end{center}
\end{figure}

\begin{figure}
[ptb]
\begin{center}
\includegraphics[
height=2.1932in,
width=4.299in
]%
{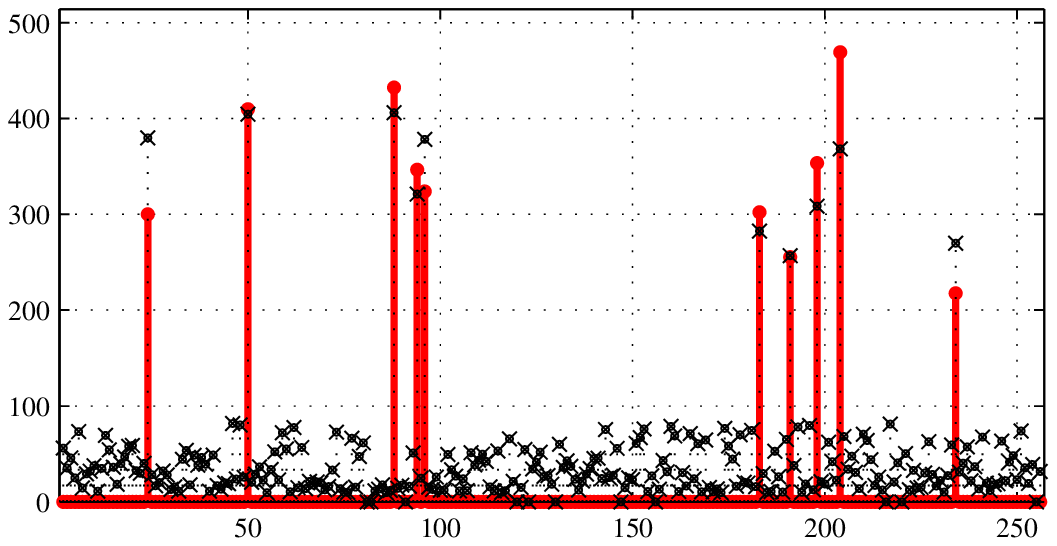}%
\caption{Single realization reconstruction of $K=10$ largest signal components
of a nonsparse noisy signal.}%
\label{input_noise_stat_snr_noise_k10}%
\end{center}
\end{figure}

\textit{Example:} Consider a nonsparse signal
\begin{align*}
x(n)  & =e^{j2\pi k_{1}n/N}+0.8e^{j2\pi k_{2}n/N}+0.77e^{j2\pi k_{3}%
n/N}+0.75e^{j2\pi k_{4}n/N}\\
&  +\sum_{i=0}^{250}(\frac{1}{3})^{1+i/50}e^{j2\pi k_{i+5}n/N}%
\end{align*}
where $k_{i},$ $i=1,2,...,251+4$ are random frequency indices from $0$ to
$N-1$. Using $N=257$ and $M=192,$ the first $K=4$ components of signal are
reconstructed. The remaining $251$ signal components are considered as
disturbance. Reconstruction of $K=4$ largest components is done in $100$
independent realizations with different frequencies and positions of available
samples. The result for $SNR$ in the noise free case, obtained statistically and by using the Theorem, is%
\begin{align*}
SNR_{stat}  &  =10\log\left(  \frac{\left\Vert \mathbf{X}_{K}\right\Vert
_{2}^{2}}{\left\Vert \mathbf{X}_{K}\mathbf{-X}_{R}\right\Vert _{2}^{2}%
}\right)  =23.1476\\
SNR_{theor}  &  =10\log\left(  \frac{\left\Vert \mathbf{X}_{K}\right\Vert
_{2}^{2}}{K\frac{N-M}{M}\left\Vert \mathbf{X-X}_{K}\right\Vert _{2}^{2}%
}\right)  =23.1235.
\end{align*}
Note that the calculation of $SNR_{theor}$ is simple since we assumed that the
amplitudes of disturbing components are coefficients of a geometric series.
One realization with $K=4$ is presented in
Fig. \ref{input_noise_stat_snr_noise}. The case when $K=10$ is presented in
Fig. \ref{input_noise_stat_snr_noise_k10}. Red signal (with dots) represents the reconstructed 
signal with assumed sparsity and the signal with black crosses represents the original nonsparse signal.%

%

In the case of additive complex-valued noise of variance $\sigma_{\varepsilon
}^{2}=2$ the results are
\begin{align*}
SNR_{stat}  &  =10\log\left(  \frac{\left\Vert \mathbf{X}_{K}\right\Vert
_{2}^{2}}{\left\Vert \mathbf{X}_{K}\mathbf{-X}_{R}\right\Vert _{2}^{2}%
}\right)  =17.0593\\
SNR_{theor}  &  =10\log\left(  \frac{\left\Vert \mathbf{X}_{K}\right\Vert
_{2}^{2}}{K\frac{N-M}{M}\left\Vert \mathbf{X-X}_{K}\right\Vert _{2}^{2}%
+\frac{K}{M}N\sigma_{\varepsilon}^{2}}\right)  =17.0384.
\end{align*}
The decrease in the SNR due to noise is
\[
\Delta SNR_{theor}=10\log\left(  \frac{K\frac{N-M}{M}\left\Vert \mathbf{X-X}%
_{K}\right\Vert _{2}^{2}}{K\frac{N-M}{M}\left\Vert \mathbf{X-X}_{K}\right\Vert
_{2}^{2}+\frac{K}{M}N\sigma_{\varepsilon}^{2}}\right)  =-6.0851.
\]
The simulation is repeated with $M=128$ and the same noise. The SNR values are
$SNR_{theor}=14.3345$ and $SNR_{stat}=14.4980.$

\section{Conclusions}

The goal of compressive sensing \ is to reconstruct a sparse signal using a
reduced set of available samples.\ It can be done by minimizing the sparsity
measure and available samples. A simple algorithm for signal reconstruction is
presented. One step reconstruction and an iterative procedure of the
reconstruction algorithm are given. Noisy environment is taken into account as
well. The input noise can degrade the reconstruction limit. However, as far as
the reconstruction is possible, the noise caused by missing samples manifests
its influence to the results accuracy in simple and direct way through the
number of missing samples and signal sparsity. The accuracy of the final
result is related to the input noise intensity, number of available samples
and the signal sparsity. A theorem presenting error in the case when the
reconstruction algorithm defined for reconstruction of sparse signals are used
in for nonsparse signals reconstruction is defined as well. The theory is
checked and illustrated on numerical examples.

\section{Appendix: Reconstruction Conditions}

Consider an $N$-dimensional vector $\mathbf{X}$ whose sparsity is $K$ and its
$M$ measurements $\mathbf{y=AX}$, where the measurements matrix $\mathbf{A}$
is an $M\times N$ matrix, with $K<M\leq N$. A reconstruction vector
$\mathbf{X}$ can be achieved from a reduced set of samples/measurements using
the sparsity measures minimization.

The $\ell_{1}$-norm based solution of constrained sparsity measure
minimization
\begin{equation}
\min\left\Vert \mathbf{X}\right\Vert _{1}\text{ \ \ subject to\ \ }%
\mathbf{y=AX} \label{minL1}%
\end{equation}
is the same as the $\ell_{0}$-norm based solution of
\[
\min\left\Vert \mathbf{X}\right\Vert _{0}\text{ \ \ subject to \ }%
\mathbf{y=AX}%
\]
if the measurements matrix $\mathbf{A}$ satisfies the restricted isometry
property for a $2K$ sparse vector%
\[
\left(  1-\delta_{2K}\right)  \left\Vert \mathbf{X}_{2K}\right\Vert _{2}%
^{2}\leq\frac{1}{E_{\Psi}}\left\Vert \mathbf{A}_{2K}\mathbf{X}_{2K}\right\Vert
_{2}^{2}\leq\left(  1+\delta_{2K}\right)  \left\Vert \mathbf{X}_{2K}%
\right\Vert _{2}^{2}%
\]
with a sufficiently small $\delta_{2K}$. Constant $E_{\Psi}$ is the energy of
columns of measurement matrix $\mathbf{A}$. For normalized energy $E_{\Psi}%
=1$, while for the measurement matrix obtained using $M$ rows of the standard
DFT matrix $E_{\Psi}=M$. If the signal $\mathbf{X}$ is not sparse then the
solution of minimization problem (\ref{minL1}) denoted by $\mathbf{X}_{R}$
will satisfy
\begin{equation}
\left\Vert \mathbf{X}_{R}\mathbf{-X}\right\Vert _{2}\leq C_{0}\frac{\left\Vert
\mathbf{X}_{K}\mathbf{-X}\right\Vert _{1}}{\sqrt{K}} \label{BoundL1R}%
\end{equation}
where $\mathbf{X}_{K}$ is $K$ sparse signal corresponding to $K$ largest
values of $\mathbf{X}$. If the signal $\mathbf{X}$ is of sparsity $K$ then
$\left\Vert \mathbf{X}_{K}\mathbf{-X}\right\Vert _{2}=0$ and $\mathbf{X}%
_{R}\mathbf{=X}$. In the case of noisy measurements when
\[
\left\Vert \mathbf{y-AX}\right\Vert _{2}\leq\epsilon
\]
then \cite{EW}
\[
\left\Vert \mathbf{X}_{R}\mathbf{-X}\right\Vert _{2}\leq C_{0}\frac{\left\Vert
\mathbf{X}_{K}\mathbf{-X}\right\Vert _{1}}{\sqrt{K}}+C_{1}\epsilon
\]
where $C_{0}$ and $C_{1}$ are constants depending on $\delta_{2K}$. For
example, with $\delta_{2K}=1/4$ constants are $C_{0}\leq5.5$ and $C_{1}\leq6$,
\cite{EW}.

\end{document}